\newcommand{\beq}{\begin{equation}}
\newcommand{\eeq}{\end{equation}}
\newcommand{\bea}{\begin{eqnarray}}
\newcommand{\eea}{\end{eqnarray}}
\newcommand{\bear}{\begin{eqnarray*}}
\newcommand{\eear}{\end{eqnarray*}}

\documentclass[12pt]{article}
\begin{document}

\title{Generalization of the matrix product ansatz for integrable chains}
\author{F.~C.~Alcaraz, M.~J.~Lazo\\
\small \it 
Instituto de F\'{\i}sica de S\~ao Carlos, 
Universidade de S\~ao Paulo,\\
\small \it  C.P. 369,13560-970, S\~ao Carlos, SP, Brazil }
\date{}

\maketitle

\begin{abstract}
We present a  general formulation of the matrix product ansatz for exactly 
integrable chains on periodic lattices. This new formulation extends the 
matrix 
product ansatz  present on our previous articles ( F. C. Alcaraz and M. J. 
Lazo 
{\it J. Phys. A: Math. Gen.} {\bf 37} (2004) L1-L7 and {\it J. Phys. A: Math. 
Gen.} 
{\bf 37} (2004) 4149-4182.)
\end{abstract}

\vskip 1em


In \cite{alclazo1} (to which we refer  hereafter as I) and \cite{alclazo2},
  we formulate a matrix product 
ansatz (MPA) for a large family of exactly integrable spin chains such as the 
anisotropic 
Heisenberg model,  Fattev-Zamolodchikov model, Izergin-Korepin model,  
Sutherland model, t-J model, Hubbard model, etc. In this note we present a 
generalization 
of the MPA for periodic quantum chains. The generalization is important since 
it allows, at least in some cases, finite-dimension representations of the 
matrices defining the MPA. 
In order to illustrate this generalization we consider the standard XXZ 
quantum chain with 
periodic boundary condition, 
\begin{equation}
\label{a1}
H = -\frac{1}{2}\sum_{i=1}^L(
\sigma_i^x\sigma_{i+1}^x +
\sigma_i^y\sigma_{i+1}^y +
\Delta \sigma_i^z\sigma_{i+1}^z ),
\end{equation}
where 
$\sigma_i^x,\sigma_i^y,\sigma_i^z$ are spin-$\frac{1}{2}$ Pauli matrices 
located at the 
$L$ sites of the chain. An arbitrary eigenstate of (\ref{a1}) $|\psi_{n,p}>$,  
in the sector with a 
number $n$ of up spins ($n=0,1,\ldots$) and momentum $p = \frac{2\pi}{L}j$,  
($j=0,1,\ldots,L-1$) is given by 
\begin{equation}
\label{a2}
|\psi_{n,p}> = \sum_{1\leq x_1<x_2<\cdots <x_n\leq L} f(x_1,\ldots,x_n) 
|x_1,\ldots,x_n>,
\end{equation}
where $|x_1,\ldots,x_n>$ denotes the coordinates of the up spins of an 
arbitrary configuration.

As in I we make a one-to-one correspondence between the configurations of 
spins and product 
of matrices. The matrix product associated to a given configuration is 
obtained by associating 
to the sites with down and up spins a matrix $E$ and $A$, respectively. The 
unknown amplitudes in 
(\ref{a2}) are obtained by associating them to the MPA
\begin{equation}
\label{a3}
f(x_1,\ldots,x_n) \Leftrightarrow 
E^{x_1-1}AE^{x_2-x_1-1}A\cdots E^{x_n-x_{n-1}-1}AE^{L-x_n}.
\end{equation}
Actually $E$ and $A$ are abstract operators with an associative product. 
A well defined eigenfunction is obtained, apart from a normalization factor, 
if all the amplitudes are related uniquely, due to the algebraic relations (to be fixed) 
among the matrices $A$ and $E$.
 Equivalently the correspondence (\ref{a3}) implies that, in the subset 
of words (products of matrices) 
of the algebra containing $n$ matrices $A$ and $L-n$ 
matrices $E$ there exists only a single independent word ("normalization constant"). 
The relation 
between any two words is a $c$ number that gives the ratio between the corresponding amplitudes 
in (\ref{a3}). 

We could  also formulate the ansatz (\ref{a3}) by associating a complex 
number to the single independent word. We can choose any operation on the 
matrix products that gives a non-zero scalar. In  the original formulation of 
the MPA  
with periodic boundary conditions \cite{alclazo1,alclazo2}
  the trace operation was chosen to produce this scalar

\begin{equation}
\label{a4}
f(x_1,\ldots,x_n) =
\mbox{Tr}
[E^{x_1-1}AE^{x_2-x_1-1}A\cdots 
E^{x_n-x_{n-1}-1}AE^{L-x_n}\Omega_p].
\end{equation}
The matrix $\Omega_p$ was chosen to have a given  algebraic 
relation with the matrices $E$ and $A$.  Recently, Golinelli and 
Mallick \cite{golimali} have shown that in the particular case of the 
asymmetric exclusion problem in a periodic chain it is possible to 
formulate the ansatz only by imposing relations between the matrices 
$E$ and $\Omega_p$. The relations between $A$ and $\Omega_p$ being totally 
arbitrary. Actually, as we are going to show in this note, we do not need 
to impose any algebraic relation between the matrices $E$ and $A$ with 
$\Omega_p$. The matrix $\Omega_p$ can be  just any arbitrary matrix 
that produces 
a non vanishing trace  in (\ref{a4}). This observation is not particular 
for the present model. It is valid for any of the exactly integrable quantum 
chains solved in the original formulation of the MPA 
 presented in I. Instead of restricting the MPA 
 with the 
trace operation, as in I, 
we  consider the more general formulation   (\ref{a3}). 

Since the eigenfunctions produced by the ansatz have a well defined momentum,  
$p = \frac{2\pi}{L}j$ ($j=0,\ldots,L-1$), the correspondence (\ref{a3}) implies   
 the following constraints  for the matrix products 
appearing in the ansatz (\ref{a3})
\begin{equation}
\label{a5}
E^{x_1-1}AE^{x_2 -x_1 -1} \cdots A 
E^{L-x_n}  =   
  e^{-ip} E^{x_1}AE^{x_2-x_1-1}\cdots A 
E^{L-x_n-1},
\end{equation}
for $x_n \leq L-1$, and for $x_n =L$ 
\begin{equation}
\label{a6}
E^{x_1-1}AE^{x_2 -x_1 -1} \cdots A 
  =   
  e^{-ip}AE^{x_1-1}A\cdots A 
E^{L-x_{n-1}-1}.
\end{equation}

The eigenvalue equation 
\begin{equation}
\label{I8}
H |\psi_{n,p}> = e|\psi_{n,p}>,
\end{equation}
gives us relations among the amplitudes $f(x_1,\ldots,x_n)$ defining the eigenfunctions $|\psi_{n,p}>$. 
As a consequence of the correspondence (\ref{a3}) these relations give 
 two types of 
constraints for the algebraic relations of the matrices $A$ and $E$.  The 
first type of relations
come from the  
configurations where all the up spins are at distances larger than the unity. 
The algebraic  relations coming from  these relations 
 are solved by identifying the matrix $A$ as composed by $n$-spectral 
dependent 
matrices, as in (I.27), 
\begin{equation}
\label{a7} 
A = \sum _{j=1}^n A_{k_j}E,
\end{equation}
where the matrices $A_{k_j}$ obey the commutations relations 
\begin{equation}
\label{a8} 
EA_{k_j} = e^{ik_j}A_{k_j}  E.
\end{equation}
The relations (\ref{a7}) and (\ref{a8}),  applied to the  algebraic constraints implied by the 
eigenvalue equation 
(\ref{I8}) and to 
(\ref{a5}), give us the energy $e$ and momentum $p$ as a function of the spectral 
parameters 
\begin{equation}
\label{a9}
e= \frac{\Delta}{2}(4n-L) - 2\sum_{j=1}^n\cos k_j,\;\;\;\; p = \sum_{j=1}^n 
k_j.
\end{equation}
The second type of relations, comming from the amplitudes where the spins are 
at 
nearest-neighbour positions, imply the commuting relations among the matrices 
$A_{k_j}$:
\begin{equation}
\label{a10}
A_{k_j}A_{k_l} = s(k_j,k_l) A_{k_l}A_{k_j}, 
\end{equation}
where 
\begin{equation}
\label{a11}
s(k_j,k_l) = - \frac{1+e^{i(k_j+k_l)} -2\Delta e^{ik_j}}{1+e^{i(k_j+k_l)} 
-2\Delta e^{ik_l}}.
\end{equation}

The spectral parameters $\{k_1,\ldots,k_n\}$, free up to know, are fixed by 
using (\ref{a7}), 
(\ref{a8}) in the remaining relation (\ref{a6}), giving us
\begin{equation}
\label{a12}
e^{ik_jL} = -\prod_{l=1}^n s(k_j,k_l).
\end{equation}
The solutions of (\ref{a12}), when inserted in (\ref{a9}) give us the 
eigenenergies.
The fact that the correspondence (\ref{a3}) is exact implies that, apart from an 
overall normalization constant, any amplitude $f(x_1,\ldots,x_n)$ can be calculated 
exactly.

 In the present formulation of the MPA it is possible to produce 
finite-dimensional 
representations for the matrices $A$ and $E$ \cite{thanks}. For a given 
solution 
$\{k_1,\ldots,k_n\}$ of the spectral parameter equations (\ref{a12}), in 
the 
sector with $n$ particles, the matrices $E$ and $\{A_{k_1},\ldots,A_{k_n}\}$ 
have 
the following finite-dimensional representation
\begin{eqnarray}
\label{a13}
&&E = 
\bigotimes_{l=1}^n  \left(\matrix{1 & 0 \cr 0 &e^{-ikl}}\right), \nonumber \\
&&A_{k_j} = \left[\bigotimes_{l=1}^{j-1}\left(
\matrix{s(k_j,k_l) &0 \cr 0& 1}\right) \right]
\otimes \left( \matrix{ 0 & 1 \cr 0 & 0}\right) 
 \bigotimes_{l=j+1}^n \left( \matrix{ 1 & 0 \cr 0 & 1 } \right),
\end{eqnarray}
where $s(k_j,k_l)$ are given by (\ref{a11}) and $A$ is obtained by using 
(\ref{a7}).  The dimension of the representation is $2^n$ and the products 
appearing on the ansatz have trace zero. If we want a 
formulation of the  matrix product ansatz where the trace operation is used, 
as in the formulation (\ref{a4}), it is quite simple to produce the 
matrix $\Omega_p$ that gives a non-zero value for the trace. We should stress 
that  in the original formulation of the ansatz in \cite{alclazo1,alclazo2}, 
it was 
required unnecessary 
algebraic relations among the matrices $E$ and $A$ that probably would have 
only infinite dimensional representations.
The existence of the finite representations, in the present formulation, simplifies the 
calculation of the amplitudes.

We conclude this note by mentioning that all exact solutions presented for 
periodic quantum chains in \cite{alclazo1,alclazo2,alclazo3} can be reobtained 
by using the formulation of 
  the MPA presented in this note.

\end{document}